\documentstyle[12pt]{article}

\begin{document}
\title{\textbf{Stretching magnetic fields by dynamo plasmas in Riemannian knotted tubes}} \maketitle
{\sl {L.C. Garcia de Andrade}\newline
Departamento de F\'{\i}sica
Te\'orica-IF-Universidade do Estado do Rio de Janeiro\\[-3mm]
Rua S\~ao Francisco Xavier, 524\\[-3mm]Cep 20550-003, Maracan\~a, Rio de Janeiro, RJ, Brasil\\[-3mm]\\[-3mm]\vspace{0.1cm} \paragraph*{Recently Shukurov et al [Phys Rev E 72, 025302 (2008)], made use of non-orthogonal curvilinear coordinate system on a dynamo Moebius strip flow, to investigate the effect of stretching by a turbulent liquid sodium flow. In plasma physics, Chui and Moffatt [Proc Roy Soc A 451,609,(1995)] (CM), considered non-orthogonal coordinates to investigate knotted magnetic flux tube Riemann metric. Here it is shown that, in the unstretching knotted tubes, dynamo action cannot be supported. Turbulence there, is generated by suddenly braking of torus rotation. Here, use of CM metric, shows that stretching of magnetic knots, by ideal plasmas, may support dynamo action. Investigation on the stretching in plasma dynamos, showed that in diffusive media [Phys Plasma \textbf{15},122106,(2008)], unstretching unknotted tubes do not support fast dynamo action. Non-orthogonal coordinates in flux tubes of non-constant circular section, of positive growth rate, leads to tube shrinking to a constant value. As tube shrinks, curvature grows enhancing dynamo action. }
\newpage
\section{Introduction}
Earlier R Ricca \cite{1} has developed an interesting method of
investigating the vortex filaments by applying a very popular method
from Einstein gravity \cite{2}, called Ricci rotation coefficients to
obtain covariant or intrinsic equations equivalent to Da Rios
equation and applied to solitons in n-dimensional space.
The manifold considered is a not necessarily, torsion-free
Riemannian space endowed with a Levi-Civita connection. In this
paper, attention is drawn to a
simple analytical Riemannian geometrical model, of a metric of a
knotted magnetic flux tube, previously investigated by Chui and
Moffatt \cite{3}. In their work they considered applications to magnetic helicity and energy, but did not take into account implications of their metric for dynamo action. In the present paper one pretends to fill this gap by considered the Ricca's 3D RRCs tool to obtain an invariant classification of knotted magnetic flux tube in its unstretching and non-dynamos aspects, that allows us to associate for example, the vanishing of some of the components of RRCs and the non-existence of dynamo action. In principle the reasoning here applies to kinematic dynamos. Other covariant model important in magnetic reconnection of astrophysical plasmas has been recently developed by Titov et al \cite{4}.
The ideas of simple dynamo
models in Riemannian space were inspired by the cat
Arnold's dynamo
first toy model \cite{5} uniformly stretched in Riemannian space. More recently one has investigated the application of Vishik \cite{6} anti-fast dynamo theorem to the dynamo plasma, showing that only a slow plasma dynamo is supported when tubes are unstretched. Though most of stellarators and tokamaks are metalic unstretched tubes their dynamo magnetic flow inside maybe stretch, so in principle fast dynamos may happen in tokamaks even in Euclidean spaces. Recently Nornberg at al \cite{7} have shown that the stretch-twist and fold fast dynamo method \cite{8} could be obtained from Madison torus experiment. The paper is organized as follows: In
section II the model is presented and the unstretching condition is imposed on the magnetic field flow. Section III presents the
computations of stretching Ricci rotation coefficients and the magnetic energy of helical knotted Riemannian tube manifolds. In section IV it is shown, that in the case of orthogonal and non-orthogonal coordinates, dynamo action can be present. Conclusions are presented in section V.
\section{Riemann-Chui-Moffatt knotted magnetic flux tube metric}
This section presents the Riemann
metric proposed by Chui and Moffatt of a curvilinear coordinates
$(s,r,{\phi})$. Note that magnetic lines along the tube are computed in the Frenet frame $(\textbf{t},\textbf{n},\textbf{b})$ which obeys the following evolution equations
\begin{equation}
\frac{d\textbf{t}}{ds}={\kappa}(s)\textbf{n}
\label{1}
\end{equation}
\begin{equation}
\frac{d\textbf{n}}{ds}=-{\kappa}(s)\textbf{t}+{\tau}\textbf{b}
\label{2}
\end{equation}
\begin{equation}
\frac{d\textbf{b}}{ds}=-{\tau}(s)\textbf{n}
\label{3}
\end{equation}
where, $\textbf{t}$ is the vector tangent to magnetic lines while in the plane orthogonal to them, where the normal $\textbf{n}$ and the binormal vector $\textbf{b}$ are defined. Here ${\kappa}$ and ${\tau}$ are Frenet curvature and torsion scalars.An arbitrary vector $\textbf{x}$ in the tube $\cal{T}$ can be expressed as
\begin{equation}
\textbf{x}(s)=\textbf{X}+r(\cos{\theta}\textbf{n}+\sin{\theta}\textbf{b})
\label{4}
\end{equation}
Where $(r,{\theta})$ are polar coordinates that are defined in the plane through $\textbf{X}(s)$ defined by the vectors $(\textbf{n},\textbf{b})$. The boundaries of a ribbon surface ${C}={\theta}=constant$ along the tube, and its curve of intersection ${\Gamma}_{\theta}$ with the boundary tube ${\partial}\cal{T}$. Here ${\cal{N}}$ is the Gauss linking topological number of $(C,{\Gamma}_{\theta})$ which my be positive, negative or even zero. Now, let us define a twist angle
\begin{equation}
{\phi}={\theta}+\frac{2{\pi}\cal{N}s}{L}
\label{5}
\end{equation}
At each point $(s,r,{\phi})$, there is a unique magnetic surface
\begin{equation}
\chi={\chi}(r,s,{\phi})
\label{6}
\end{equation}
Inversion of this relation allows us to use now the new coordinate system
$(s,{\chi},{\phi})$. By definition of magnetic surface
\begin{equation}
\textbf{B}.{\nabla}\chi=0\label{7}
\end{equation}
Implies that $\textbf{B}$ has no component in the ${\nabla}{\chi}$ direction. Chui and Moffatt define a shape-function along the magnetic flux tube as $R(s,{\chi},{\phi})$ which allows us to define the tube geometrically as
\begin{equation}
\textbf{x}(s)=\textbf{X}+R(\cos{\theta}\textbf{n}\sin{\theta}\textbf{b})
\label{8}
\end{equation}
Chui-Moffatt used the following fundamental differential form
\begin{equation}
d\textbf{x}(s)=\textbf{e}_{1}ds+\textbf{e}_{2}d{\chi}+\textbf{e}_{3}d{\phi}
\label{9}
\end{equation}
Where the vector basis frame $(\textbf{e}_{1},\textbf{e}_{2},\textbf{e}_{3})$
is given by \cite{3}
\begin{equation}
\textbf{e}_{1}=(1-{\kappa}\cos{\theta})\textbf{t}+R_{s}(\cos{\theta}-R{\tau}^{*}\sin{\theta})\textbf{n}+(R_{s}\sin{\theta}+R{\tau}^{*}cos{\theta})\textbf{b}
\label{10}
\end{equation}
\begin{equation}
\textbf{e}_{2}=R_{\chi}(\cos{\theta}\textbf{n}+\sin{\theta}\textbf{b})
\label{11}
\end{equation}
\begin{equation}
\textbf{e}_{3}=(R_{\phi}\cos{\theta}-R\sin{\theta})\textbf{n}+(R\cos{\theta}+R_{\phi}\sin{\theta})\textbf{b}
\label{12}
\end{equation}
Where ${\tau}^{*}={\tau}-\frac{2{\pi}\cal{N}}{L}$ where L is the length of the tube. The Riemann non-orthogonal metric tensor $g_{ij}$ where $(i,j=1,2,3)$ is given by
\begin{equation}
g_{ij}=(\textbf{e}_{i}.\textbf{e}_{j})\label{13}
\end{equation}
which in matrix format is given by
\begin{equation}
$$\displaylines{\pmatrix{{(1-R{\kappa}\cos{\theta})^{2}+R^{2}{{\tau}^{*}}^{2}+{R_{s}}^{2}}
&{R_{\chi}R_{s}}& {R^{2}{{\tau}^{*}}^{2}+{R_{s}}R_{\phi}}\cr{R_{\chi}R_{s}}&{{R_{\chi}}^{2}}&{R_{\chi}R_{\phi}}
\cr{R^{2}{{\tau}^{*}}^{2}+{R_{s}}^{2}}&{R_{\chi}R_{\phi}R^{2}}+{{R_{\phi}}^{2}}&R^{2}+{R_{\phi}}^{2}\cr}\cr}$$
\label{14}
\end{equation}
In the next section one shall see that the poloidal and toroidal magnetic field given by
\begin{equation}
\textbf{B}_{T}=B^{1}\textbf{e}_{1}
\label{15}
\end{equation}
\begin{equation}
\textbf{B}_{P}=B^{3}\textbf{e}_{3}
\label{16}
\end{equation}
Are constrained by imposing the unstretching magnetic flux tube condition
\begin{equation}
(\textbf{B}.{\nabla})\textbf{v}=0\label{17}
\end{equation}
onto the magnetic self-induction equation
\begin{equation}
{d}_{t}\textbf{B}=(\textbf{B}.{\nabla})\textbf{v}+{\eta}{\nabla}^{2}\textbf{B}
\label{18}
\end{equation}
Here $B^{2}$ vanishes due to the magnetic surface condition above.
\section{Knot energy in unstretched flux tubes}
Just for comparison one writes here the Germano-Ricca \cite{9} Riemann metric of orthogonal coordinates in magnetic flux tube as used in our previous Riemanian flux tubes papers \cite{9} is appended
\begin{equation}
dl^{2}= dr^{2}+[r^{2}d{{\theta}_{R}}^{2}+K^{2}(s)ds^{2}] \label{19}
\end{equation}
where metric factor $K(r,s)=(1-{\kappa}rcos{\theta})$ which is contained also in the Chui-Moffatt metric, showing that this metric is rather simpler than the Chui-Moffatt one. Despite this backdraw, there is a clear advantage to use the more general knotted Riemannian metric, and this is the richness of mathematical and physical information, as one shall see in the rest of the paper. Of course, Germano-Ricca metric (\ref{17}) is very useful in the solar loops topology. The present Chui-Moffat knotted covariant metric also helps one to investigate reconnection of solar tubes by considering thin tubes. Note that by considering the index notation $(\textbf{e}_{A})$ with $(A=1,2)$ one may express the unstretching condition above as
\begin{equation}
(\textbf{B}.{\nabla})\textbf{v}=B^{A}{\partial}_{A}(v^{i}\textbf{e}_{i})= B^{A}v^{i}{\partial}_{A}(\textbf{e}_{i})=B^{A}v^{i}{{\Gamma}^{j}}_{Ai}\textbf{e}_{j}
\label{20}
\end{equation}
Note that to simplify matters one considered that the modulus of the velocity flow is constant and ${{\Gamma}^{j}}_{Ai}$ is the so-called Ricci rotation coefficients, so-commonly use also in general relativity. It is clear that these objects here may be written in terms of the Ricci coefficients written in terms of the Frenet frame as
\begin{equation}
\textbf{n}.{\partial}_{s}\textbf{t}= \textbf{n}.\kappa\textbf{n}=\kappa={\Gamma}_{nss}  \label{21}
\end{equation}
\begin{equation}
\textbf{n}.{\partial}_{s}\textbf{b}=-\tau={\Gamma}_{nsb}  \label{22}
\end{equation}
\begin{equation}
\textbf{t}.{\partial}_{s}\textbf{n}=-\kappa={\Gamma}_{ssn}  \label{23}
\end{equation}
\begin{equation}
\textbf{b}.{\partial}_{s}\textbf{n}=-\tau={\Gamma}_{bsn} \label{24}
\end{equation}
From these last four expressions one infers that the Frenet-Ricci coefficients above have the following symmetries
\begin{equation}
{\Gamma}_{jki}=-{\Gamma}_{ikj}
\label{25}
\end{equation}
As in Einstein's general relativistic spacetime, where the vanishing of Ricci rotation coefficients, imply that the spacetime is Minkowskian or flat, here the vanishing of Frenet-Ricci rotation coefficients means that the curvature and torsion of the curve vanishes, meaning that the line or magnetic fields are flat. A similar formalism to the one used in this paper was used by Ricca \cite{1} to investigate filaments and solitons. From the above expressions one may use the following Ricci-like  relation
\begin{equation}
{\partial}_{A}(\textbf{e}_{i})={{\Gamma}^{j}}_{Ai}\textbf{e}_{j}
\label{26}
\end{equation}
Which clearly shows that
\begin{equation}
{{\Gamma}^{j}}_{Ai}=0
\label{27}
\end{equation}
is equivalent to unstretching. One also notes from here, that not all Ricci rotation like coefficients ${\Gamma}$ vanish, so one may say that of course, the space does not need necessarily be flat. Next one shall consider, the specific case of Chui-Moffatt covariant metric of knotted flux tubes. In the case of the helical knotted tube, the torsion ${\tau}_{0}$ and curvature ${\kappa}_{0}$ are constants, a long but straightforward computation leads to the following stretching expressions in terms of the Ricci rotation coefficients as
\begin{equation}
\textbf{n}.(\textbf{B}.{\nabla})\textbf{v}=B^{A}v^{i}{{\Gamma}^{j}}_{Ai}\textbf{n}.\textbf{e}_{j}
\label{28}
\end{equation}
The Ricci rotation coefficients needed can be computed as
\begin{equation}
{{\Gamma}^{n}}_{s1}=\textbf{n}.{\partial}_{s}\textbf{e}_{1}={\kappa}_{0}+(2\frac{R_{s}{\pi}\cal{N}}{L}+{{\tau}^{*}}_{0}[R_{s}+\frac{R{\pi}\cal{N}}{L}])sin{\theta}+{{\tau}^{*}}_{0}[ R_{s}+\frac{R{\pi}\cal{N}}{L}]cos{\theta}
\label{29}
\end{equation}
\begin{equation}
{{\Gamma}^{n}}_{s2}=\textbf{n}.{\partial}_{s}\textbf{e}_{2}=[-R_{{\chi}s}\cos{\theta}-R_{\chi}[\frac{2{\pi}\cal{N}}{L}+{{\tau}^{*}}_{0}]\cos{\theta}
\label{30}
\end{equation}
\begin{equation}
{{\Gamma}^{n}}_{s3}=\textbf{n}.{\partial}_{s}\textbf{e}_{3}=[R_{{\phi}s}+R(\frac{2{\pi}\cal{N}}{L}-{{\tau}^{*}}_{0}]\cos{\theta}-R_{\phi}(\frac{2{\pi}\cal{N}}{L}+{{\tau}^{*}}_{0})+R_{s})\sin{\theta}
\label{31}
\end{equation}
\begin{equation}
{{\Gamma}^{s}}_{s2}=\textbf{t}.{\partial}_{s}\textbf{e}_{2}=-R_{\chi}{\kappa}\cos{\theta}
\label{32}
\end{equation}
This last expression has two physically interesting interpretations. Since either, the curvature ${\kappa}$ vanishes, which is the trivial straight flux tube, or $R_{\chi}$ vanishes, which is the case of total constant cross-section tube. Let us now compute the knot magnetic energy of the unstretched constant cross-section flux tube, by noting that the RRC could be expressed in terms of the Frenet frame directly, instead of the above triad. This allows us to write
\begin{equation}
{{\Gamma}}_{i1j}=\textbf{e}_{i}.{\partial}_{s}\textbf{e}_{j}=
{\gamma}_{i{\alpha}}\textbf{E}^{\alpha}{\partial}_{s}[{\gamma}_{j{\beta}}\textbf{E}^{\beta}]
\label{33}
\end{equation}
where the vector $\textbf{E}_{\alpha}=(\textbf{t},\textbf{n},\textbf{b})$ is the compact notation for the Frenet frame. Explicitly this equation yields
\begin{equation}
{{\Gamma}}_{112}=\frac{1}{2}{\partial}_{s}[{{\gamma}_{11}}^{2}+{{\gamma}_{12}}^{2}+{{\gamma}_{13}}^{2}]
\label{34}
\end{equation}
The Ricci coefficients ${\gamma}_{ij}$ can be determined from the values in the expressions for the triad frame $\textbf{e}_{j}$ above. The non-vanishing components of these coefficients are
\begin{equation}
{{\gamma}}_{11}=K(r,s)
\label{35}
\end{equation}
\begin{equation}
{{\gamma}}_{12}=-R{\tau}^{*}\sin{\theta}
\label{36}
\end{equation}
\begin{equation}
{{\gamma}}_{13}=R{\tau}^{*}\cos{\theta}
\label{37}
\end{equation}
\begin{equation}
{{\gamma}}_{22}=R_{\chi}\cos{\theta}
\label{38}
\end{equation}
\begin{equation}
{{\gamma}}_{23}=R_{\chi}\sin{\theta}
\label{39}
\end{equation}
\begin{equation}
{{\gamma}}_{32}=-R sin{\theta}
\label{40}
\end{equation}
\begin{equation}
{{\gamma}}_{33}=R cos{\theta}\label{41}
\end{equation}
where in these expressions one already takes into account that $R_{\phi}$ and $R_{s}$ vanish. This means that the tube is uniform and its cross-section is constant. Substitution of these coefficients into (\ref{33}) yields
\begin{equation}
{{\Gamma}}_{112}=\frac{1}{2}[K^{2}+{{\tau}}^{2}R^{2}]
\label{42}
\end{equation}
Next one shall see that these affine connections are very useful to compute the ratio between toroidal and poloidal magnetic fields. This can be seen , since according to the unstretching condition above
\begin{equation}
B^{A}{{\Gamma}}_{iAj}=0
\label{43}
\end{equation}
is equivalent to the equations
\begin{equation}
B^{1}{{\Gamma}}_{i1j}+
B^{3}{{\Gamma}}_{i3j}=0
\label{44}
\end{equation}
which in the first case reads
\begin{equation}
B^{1}{{\Gamma}}_{112}+
B^{3}{{\Gamma}}_{132}=0
\label{45}
\end{equation}
Now computation of the RRC  ${{\Gamma}}_{132}$ as
\begin{equation}
{{\Gamma}}_{132}=RR_{\chi}{\tau}^{*}
\label{46}
\end{equation}
yields the following expression for the ratio between toroidal and poloidal magnetic field as
\begin{equation}
\frac{B^{1}}{B^{3}}=-b:=-\frac{{\Gamma}_{112}}{{{\Gamma}}_{132}}\approx{-2RR_{\chi}{\tau}^{*}}
\label{47}
\end{equation}
Note the this expression shows us that, in the next section example, where the effective torsion ${\tau}^{*}$ vanishes, in this unstretching case , dynamo action would not be present since the toroidal component, $B^{1}$ would vanish. But note that in the next section the flux tube is stretching, so as one shall show the dynamo action can be supported in the knotted flux tube. Here one used the approximation of this tubes and helical weak effective torsion ${\tau}^{*}$. These expressions allows us immeadiatly to write the expression for the knot energy in the Riemannian flux tube manifold\cite{9}. Since the tube is unstretched one may consider that its length L is constant. Thus
\begin{equation}
M=\frac{1}{2}\int{\sqrt{g}[g_{11}b^{2}+g_{33}-2bg_{13}]ds d{\chi} d{\phi}}
\label{48}
\end{equation}
Since $R=R(\chi)$ here, the magnetic energy of the knotted MFT becomes
\begin{equation}
<M>=\frac{1}{2}{\epsilon}^{3}L^{2}V({\chi})<{(B^{3})}^{2}>
\label{49}
\end{equation}
where ${\epsilon}=(\frac{V_{\cal{T}}}{{\pi}L^{3}})^{3}$ and ${\chi}=\frac{V_{\cal{T}}}{V}$. Here $V_{\cal{T}}$ is the volume inside the whole tube while $V_{\chi}$ is the volume under the magnetic surface ${\chi}=constant$. Actually $<M>$ is an averaged or mean field energy, and we assume that the average value of $B^{3}$ squared is constant over an averaged volume. Here also
\begin{equation}
{V({\chi})}=\int{\sqrt{g}ds d{\chi} d{\phi}}
\label{50}
\end{equation}
with the above approximations we were left also with the fact that $g_{13}$ component vanishes which also facilitates our computations. Since at the magnetic surface the volume is constant from the knot energy, one must conclude that this leads to a marginal dynamo action with a constant energy.
\newpage
\section{Dynamo action in flux tubes with circular cross-section}
The first matematically analyzed fast dynamo model, as a smooth flow of uniform stretching, was the suspension of the cat map by Arnold, Zeldovich, Ruzmaikin, and Sokoloff \cite{5} in 1981. More recently, Oseledets \cite{13,14} has been investigated these dynamos in random smooth maps, following Baxendale and Rozovskii \cite{15} work. More recently, an interesting example of a two-dimensional turbulent dynamo on a Moebius strip flow was given by Shukurov, Stepanov and Sokoloff \cite{16}, with applications in liquid sodium torus dynamo experiment. Chicone, Latushkin and Smith \cite{17} have provide a theorem, where fast dynamo action could be only be supported in a two-dimensional surface, if this possess a constant negative Riemannian curvature. In this paper another new application of this theorem, is presented by showing that a fast dynamo action, can be supported in inflexionary \cite{18} twisted magnetic flux tube surface of in regions of negative curvature. This example is similar to chaotic dynamos obtained previously by Boozer \cite{19} and Thiffeault and Boozer \cite{20} who investigate eigenvalues, given by kinematic magnetic dynamo Lyapunov exponents, instead of Lefschetz numbers. More recently Vishik \cite{21} a have presented and proved a anti-fast dynamo theorem, which could be considered as a theorem for positive two-dimensional curvature as well. Recently Garcia de Andrade \cite{11} has applied Vishik's theorem in dynamo plasmas. This study is important not only to kinematic dynamo, but also to hydromagnetic dynamos \cite{22} where Lorentz force, back-reaction helps stretching dynamos by plasma flows.  In this paper, to simplify matters, the resonant condition between poloidal and toroidal frequencies. This condition could be tested experimentally in the Perm russian dynamo torus , where the frequencies profiles could be adjusted. Riemannian negative curvature of Anosov \cite{23} type appears here, naturally in inflexionary flux tube surfaces, since the scalar principal toroidal curvature of the constant cross-section, is positive, while the curvature in the toroidal eigendirection is negative in the deeper regions, where inflexion of the tube takes place. Three-dimensional conformal dynamos, \cite{24} have been obtained earlier by stretching of ideal plasma in flux tubes. But due to Zeldovich et al \cite{5} and Cowling \cite{10} anti-dynamo theorems for special symmetries of the flow, two-dimensional dynamo action in incompressible flows as dealt here, are more complex. and deserve further investigation. Here resistivity ${\epsilon}$ does not vanish over the two-dimensional Riemannian surface of the flux tubes. The importance of investigating the geometry and topology of solar flux tube dynamos, as here, stems from the work of M Schuessler \cite{25} on magnetic flux tube dynamos in solar and stellar plasma physics. Recently Ashgari-Targhi and Berger \cite{26} have investigated in detail the topology of stretch-twist and fold dynamos \cite{27} in flux tubes as well. Let us now consider the above CM metric in the case of the circular cross-section, not necessarily constant. In this case $R_{\phi}$ vanishes and the effective torsion is also assumed to vanish, which simplifies computations. Another assumption is that the modulus of the plasma flow velocity coincides as $v^{1}$ and $v^{3}$ are constants. From the expressions of the last section, one is left as
\begin{equation}
$$\displaylines{\pmatrix{(1-R{\kappa}\cos{\theta})^{2}
&R_{\chi}R_{s}& 0\cr{R_{\chi}R_{s}}&{{R_{\chi}}^{2}}&{0}
\cr{0}&{0}&R^{2}\cr}\cr}$$
\label{51}
\end{equation}
where now
\begin{equation}
\textbf{e}_{1}=(1-{\kappa}\cos{\theta})\textbf{t}+R_{s}\textbf{e}_{r}
\label{52}
\end{equation}
\begin{equation}
\textbf{e}_{3}=R[-\sin{\theta}\textbf{n}+\cos{\theta}]\textbf{b}=R\textbf{e}_{\theta}
\label{53}
\end{equation}
where now one has used the relation between the non-orthogonal $(\textbf{e}_{1},\textbf{e}_{2},\textbf{e}_{3})$ and orthogonal basis $(\textbf{t},\textbf{e}_{r},\textbf{e}_{\theta})$ given by
\begin{equation}
\textbf{e}_{\theta}=-\sin{\theta}\textbf{n}+\cos{\theta}
\label{54}
\end{equation}
and
\begin{equation}
\textbf{e}_{r}=\cos{\theta}\textbf{n}+\sin{\theta}
\label{55}
\end{equation}
 Actually, the assumption of circular cross section, could be added the assumption that the tube was axially uniform, but since an excess of symmetries can spoil dynamo action, here to simplify matters one keep up with the simplest non-orthogonal metric. When both assumptions are kept, not only $R_{s}$ but also $R_{\phi}$ vanishes. One must note that now the radial $B^{r}$ component of the magnetic field vanishes. In this case the flux tube basis is orthogonal. Besides being stretched by the plasma flow, the tubes one has  assumed above that one has a vanishing effective torsion ${\tau}^{*}$, implies that
\begin{equation}
{\tau}_{0}=\frac{2{\pi}\cal{N}}{L}\label{56}
\end{equation}
which, of course does not mean that the torsion of the tube vanishes, but that it is constant. Note also that, now torsion depends directly upon the Gauss linking topological number. From the above relation between ${\theta}$ and s, one is able to obtain the following relation between derivative operators
\begin{equation}
{{\tau}_{0}}^{-1}{\partial}_{s}=-\frac{{\omega}_{s}}{{\kappa}_{R}}{\partial}_{\theta}\label{57}
\end{equation}
where ${\kappa}_{0}=\frac{1}{r_{0}}$ is the internal curvature of the tube, which here is positive, since is circular. Here $r_{0}$ is the external radius. The expression ${\kappa}_{R}=\frac{1}{R}$, represents the internal curvature of the tube.
An easy computation shows that the relation between the poloidal and toroidal frequencies is
\begin{equation}
{\omega}_{\theta}=-{\tau}_{0}R{\omega}_{s}\label{58}
\end{equation}
Note that the resonance hypothesis ${\omega}_{\theta}={\omega}_{s}={\omega}_{0}$ implies that
\begin{equation}
{\tau}_{0}=-\frac{1}{R}\label{59}
\end{equation}
which tremendously simplify the following computations of the chaotic dynamo equation
\begin{equation}
{\partial}_{t}\textbf{B}={\nabla}{\times}({\textbf{v}}{\times}{\textbf{B}})\label{60}
\end{equation}
where diffusive Laplacian term has been suppressed by assuming that, magnetic Reynolds number is very high like in ideal astrophysical plasmas environment. By making use of the orthogonality between covariant and contravariant basis, given by
\begin{equation}
\textbf{e}^{i}.\textbf{e}_{j}={{\delta}^{i}}_{j}
\label{61}
\end{equation}
and the relations
\begin{equation}
{\partial}_{\phi}\textbf{e}_{1}={\partial}_{\phi}\textbf{e}_{3}
\label{62}
\end{equation}
\begin{equation}
{\partial}_{s}\textbf{e}_{1}=R_{ss}\textbf{e}_{r}
\label{63}
\end{equation}
and 
\begin{equation}
{\partial}_{s}\textbf{e}_{3}=R_{s}\textbf{e}_{\theta}-{\kappa}_{0}Rsin{\theta}\textbf{t}
\label{64}
\end{equation}
the first term on the LHS of the induction equation reads
\begin{equation}
{\partial}_{t}\textbf{B}={\lambda}{\textbf{B}}+v^{1}[R B^{1}R_{ss}\textbf{e}_{r}+B^{3}(R_{s}\textbf{e}_{\theta}-R{\kappa}_{0}sin{\theta}\textbf{t})]
\label{65}
\end{equation}
where one has used the fact that here, the curvature and torsion are constants and coincident $({\kappa}_{0}={\tau}_{0})$, as in helical tubes. Since the RHS of the kinematic dynamo equation above can be written as
\begin{equation}
{\nabla}{\times}(\textbf{v}{\times}\textbf{B})=(\textbf{B}.{\nabla})\textbf{v}-(\textbf{v}.{\nabla})\textbf{B}
\label{66}
\end{equation}
for divergence-free vector fields, one is able to express the scalar equations which comes from the induction equation as
\begin{equation}
[{R}_{ss}(v^{1}+v^{3})+{\lambda}R_{s}]B^{1}=[B^{3}v^{1}-v^{3}B^{1}]R_{ss}
\label{67}
\end{equation}
\begin{equation}
[{R}_{s}+\frac{{\lambda}}{v^{1}}R]B^{3}+{\partial}_{s}B^{3}=0
\label{68}
\end{equation}
\begin{equation}
{{\lambda}}(1-R{\kappa}_{0}cos{\theta})B^{1}-{\kappa}_{0}B^{3}=0
\label{69}
\end{equation}
To obtain these equations one assumed for simplicity, that the toroidal and poloidal flows are equal in modulus. Considering that, there is an equipartion, between the components of the magnetic field and $B^{1}=B^{3}$,what is commonly found in plasma physics, one is able to reduce these equations to the following system
\begin{equation}
2{R}_{ss}v^{3}+{\lambda}R_{s}=0
\label{70}
\end{equation}
\begin{equation}
{{\lambda}}(1-R{\kappa}_{0}cos{\theta})-{\kappa}_{0}=0
\label{71}
\end{equation}
which yields the following solutions
\begin{equation}
B^{3}=B_{0}e^{{\lambda}[t-\frac{\int{R(s)ds}}{v^{1}}]}
\label{72}
\end{equation}
and
\begin{equation}
R_{s}=A_{0}e^{-\frac{{\lambda}}{v^{1}}s}
\label{73}
\end{equation}
Note from last equation, that for dynamo action growth rate $({\lambda}>0)$, the integration constant has to be negative, otherwise the radius would be negative, thus one must concludes that the tube shrinks to a constant value. Mathematically this can be expressed as ${R}_{s}<0$. When the tube reaches a constant radius, being suitable to a flux tube approximation, the expression (\ref{72}) for the magnetic field reads
\begin{equation}
B^{3}=B_{0}e^{{\lambda}[t-\frac{R_{0}s}{v^{1}}]}
\label{73}
\end{equation}
A simple physical inspection of this expression shows that the magnetic field grows in time, and when the toroidal component of the plasma flow increases, due to previous tube shrinking, the magnetic field grows faster than the grow in distance can slow down his action. Inclusion of diffusion and back reaction Lorentz forces in the plasma non-ideal flow, would certainly change this scenario of a chaotic dynamo.
\section{Conclusions}
With the aid of a mathematical formalism popular in Einstein general relativity, called Ricci Rotation Coefficients, one imposes a unstretching of magnetic field lines along the knotted tube and obtains constraints which allows us to obtain a dynamo action, when one moves from one ergodic magnetic surface to another. For uniformly stretched, constant cross-section tubes, one obtains a marginal dynamo action, which generalizes the existence of non-fast dynamo plasmas for the knotted Riemannian flux tube manifold in comparison with the unkotted one given previously in the literature. The formalism applied to the dynamo action here can be also used in the investigation of anti-dynamo theorems \cite{10} and further investigation in helicity in dynamo plasmas \cite{26} can be further generalized to knotted tubes. More complicate knot energy \cite{12} magnetic integrals maybe obtained in Riemannian manifolds for more general knotted MFT. Stretching knotted magnetic flux, for uniformly stretched, non-constant cross-section tubes, is shown to lead to a simple way to obtain chaotic diffusive-free dynamos on an ideal plasma in non-orthogonal. Another example of the use of non-orthogonal coordinates in dynamo theory is on general relativistic dynamos in rotating stellar objects like torus around black hole plasmas. A detailed investigation of these ideas may appear elsewhere \cite{28}. To the interested reader, may found more details about the applications of non-orthogonal cooordinates in plasmas may be found in the book of D'haseleer et al \cite{29}. Finally, after I finish this paper I became aware of a very recent paper by Maggione et al \cite{30}, where they also address the CM Riemann metric in magnetic knots, which are defined in terms of the magnetic induction equation in ideal plasmas used above. Though the authors do not address the issue of dynamos, since their work is more complete and mathematical than ours, would be interesting to extend their work, to allow for the presence of dynamo action in magnetic knots. 
\section{Acknowledgements}
I appreciate financial  supports from UERJ and CNPq.

  \end{document}